\documentclass[12pt,a4paper]{article}
\usepackage{type1cm,amsmath,hangcaption}
\usepackage[psamsfonts]{amssymb}
\usepackage{cite}
\numberwithin{equation}{section}

\newcommand{\br}{\mathbb R}

\newcommand{\brp}{{\mathbb R} {\mathbb P}^2}

\newcommand{\ket}[1]{{|#1\rangle}{}}
\newcommand{\bra}[1]{{\langle#1|}}

\newcommand{\nn}{\nonumber\\}

\DeclareMathOperator*{\im}{{\rm Im}}

\begin{document}
%%% Title page %%%%%
\begin{titlepage}
 
 \renewcommand{\thefootnote}{\fnsymbol{footnote}}
\begin{flushright}
 \begin{tabular}{l}
 UT-02-57\\
  hep-th/0210305\\
 \end{tabular}
\end{flushright}

 \vfill
 \begin{center}
 \font\titlerm=cmr10 scaled\magstep4
 \font\titlei=cmmi10 scaled\magstep4
 \font\titleis=cmmi7 scaled\magstep4
 \centerline{\titlerm Liouville Field Theory on an Unoriented Surface} 
 \vskip 2.5 truecm

\noindent{ \large Yasuaki Hikida}\footnote{
E-mail: hikida@hep-th.phys.s.u-tokyo.ac.jp}
\bigskip

 \vskip .6 truecm
 {
 {\it Department of Physics,  Faculty of Science, University of Tokyo \\
  Hongo 7-3-1, Bunkyo-ku, Tokyo 113-0033, Japan} 
 }
 \vskip .4 truecm

 \end{center}

 \vfill
\vskip 0.5 truecm

\begin{abstract}

Liouville field theory on an unoriented surface is investigated,
in particular, the one point function on a $\brp$ is calculated.
The constraint of the one point function is obtained by using the
crossing symmetry of the two point function.
There are many solutions of the constraint and
we can choose one of them by considering the modular bootstrap.

\end{abstract}
\vfill
\vskip 0.5 truecm

\setcounter{footnote}{0}
\renewcommand{\thefootnote}{\arabic{footnote}}
\end{titlepage}

\newpage

%%%%%%%%%%%%%%%%%%%%%%%%%%%%%%%%%%%%%%%%%%%%%%%%%%%%%%%%%%%%%%%%%%%%

\section{Introduction}
\label{Intoduction}

In the recent development of the string theory, the D-branes become the 
crucial objects.  
Among other things, they are used for the investigation of the string
duality and $AdS/CFT$ correspondence. 
However, it is known that there are tadpoles in 
the configurations only with D-branes and we have to introduce the
orientifolds in order to cancel the tadpoles.
Thus, it is also important to investigate the orientifolds.

The most famous example is the type I string theory, 
which can be regarded as the type IIB string theory with (1+9) dimensional
orientifold plane.
This theory is defined on unoriented worldsheets with and without
boundary. 
Although we know well the orientifolds in the flat space, 
we have little knowledge about the orientifolds in the curved
backgrounds.
Only in the case of rational conformal field theory, 
the orientifolds have been investigated
\cite{sagnotti1,sagnotti2,sagnotti3} and their geometrical pictures are
given recently \cite{oplane1,oplane2,oplane3,oplane4,pfoplane,Hori}. 

In this paper, we consider the Liouville field theory on an unoriented
surface as the simple example of non-rational case.
Liouville field theory is also interesting because it appears in several
important systems.
This theory was much investigated about ten years ago because
of the relation with the two dimensional quantum gravity. 
It is known that the Liouville field theory is dual to the
$SL(2,\br)/U(1)$ WZW model, which appears in superstring theory as an
interesting solvable case. In addition,
the $AdS_3$ string theory is resemble to the Liouville field theory,
thus it is important in a sense of the $AdS/CFT$ correspondence. 

The Liouville field theory with boundary is studied in 
\cite{FZZ,TLB,ZZ,Hos,PT,FH,ARS} and we will follow their analysis.
First we obtain the solutions of the one point function by making use
of the crossing symmetry on the two point function.
Then we determine the precise form by considering the one loop partition
function. 
The D-branes in the $AdS_3$ space are much investigated  
\cite{Stan2,Stan3,BP,PR,LOPT,GKS,HS,RR,PS,Rajaraman,LOP,PST,BBDO,Skenderis,Yamaguchi,Ponsot,Ribault}
in the similar manner.
The orientifold of the $AdS_3$ space is also constructed in
\cite{AdSoplane}, however the constraint is too weak and we
cannot determine the precise form of the one point functions\footnote{
The one point function can be determined up to overall factor 
with the help of the geometric interpretation \cite{AdSoplane}.}.
It is a better point that we can determine the exact form
in the Liouville field theory case.

The organization of this paper is as follows.
In section \ref{LFT}, we review the Liouville field theory on a sphere
and summarize our notations.
In section \ref{1pf}, the one point function on a $\brp$ is examined.
We obtain the constraints from the crossing symmetry of the two point
functions and then we solve these constraints. 
In section \ref{crosscap}, we consider the modular bootstrap and the
precise form of the one point function is determined.
The conclusion and discussions are given in section \ref{conclusion}
and the several useful formulae are summarized in appendix \ref{formulae}.

%%%%%%%%%%%%%%%%%%%%%%%%%%%%%%%%%%%%%%%%%%%%%%%%%%%%%%%%%%%%%%%%%%%

\section{Liouville Field Theory}
\label{LFT}

The Liouville Field Theory is defined by the action
\begin{equation}
 S = \frac{1}{4 \pi} \int d^2 x \sqrt{g}  
   [ g^{\mu \nu} \partial_{\mu} \phi \partial_{\nu} \phi 
   + Q R \phi + 4 \pi \mu e^{2 b \phi} ] ~,
\end{equation}
where $g$ is the metric and $R$ is the scalar curvature.
The quantity $Q=b+1/b$ is called as the background charge
and $\mu$ is called as the cosmological constant.
By setting $\mu=0$, the stress tensors are given by
\begin{equation}
 T(z) = - (\partial \phi)^2 + Q \partial^2 \phi ~,~~
 \bar{T}(\bar{z}) = - (\bar{\partial} \phi)^2 + Q \bar{\partial}^2 \phi ~,
\end{equation}
and the central charge of the theory is $c=1+6Q^2$.
The primary fields are defined as $V_{\alpha}=\exp (2 \alpha \phi (x))$
with the conformal weights $\Delta_{\alpha}= \alpha (Q-\alpha)$.
The normalizable states correspond to the operators with
$\alpha=Q/2+iP$, where we restrict $P \geq 0$ since the operators
$V_{\alpha}$ and $V_{Q-\alpha}$ are related by so called reflection
relations \cite{zamozamo}. 

We can investigate the conformal field theory by considering
the correlation functions of the fields. In principle, the multi-point
correlation functions can be calculated from the information of the two
point functions and three point functions:
\begin{align}
& \langle V_{\alpha}(x) V_{\alpha}(y) \rangle =
  \frac{D(\alpha)}{|x-y|^{4 \Delta_{\alpha}}} ~, \nn
& \langle V_{\alpha_1} (x_1) V_{\alpha_2} (x_2) V_{\alpha_3} (x_3)\rangle 
 = 
  \frac{C(\alpha_1, \alpha_2 , \alpha_3)}{
 |x_1 - x_2 |^{2 \Delta_{12}}
 |x_2 - x_3 |^{2 \Delta_{23}}
 |x_3 - x_1 |^{2 \Delta_{31}}
    } ~,
\end{align}
where we use
\begin{align}
  \Delta_{12} = 
    \Delta_{\alpha_1} +  \Delta_{\alpha_2} -  \Delta_{\alpha_3} ~,~~ 
    \Delta_{23} = 
     \Delta_{\alpha_2} +  \Delta_{\alpha_3} -  \Delta_{\alpha_1} ~,~~ 
   \Delta_{31} =  
     \Delta_{\alpha_3} +  \Delta_{\alpha_1} -  \Delta_{\alpha_2} ~.
\end{align}

These quantities can be
obtained by using the following technique.
Among the general states, there are special states
which are degenerate 
\begin{equation}
\Phi_{m,n} = e^{\left((1-m)\frac{1}{b}  + (1-n) b \right) \phi } ~,
\end{equation}
and they satisfy some differential equations.
The simplest one is given for $\Phi_{1,2}=V_{-b/2}$ as
\begin{equation}
 \left(\frac{1}{b^2} \partial^2 + T(z) \right) V_{-\frac{b}{2}} = 0 ~.
\label{diff}
\end{equation}
When considering the operator product expansions including the degenerate
states, these differential equations restrict the number of primary
fields. 
For the above example $\Phi_{1,2} = V_{-b/2}$, we find
\begin{equation}
 V_{-\frac{b}{2}} V_{\alpha} 
  \sim C_+ V_{\alpha - \frac{b}{2}} + C_- V_{\alpha + \frac{b}{2}} ~,
\label{OPE}
\end{equation}
where the coefficients can be calculated as $C_+ =1$ and
\begin{equation}
 C_- = - \mu \pi \frac{\gamma(2 b \alpha - 1 - b^2)}
                      {\gamma(-b^2) \gamma(2 b \alpha)} ~,~~
  \gamma(x) = \frac{\Gamma(x)}{\Gamma(1-x)} ~.
\end{equation}
Using this operator product expansion, we can evaluate the three
point function including this field in two ways.
Equating two quantities, we obtain the constraint and the solution is
given by
\begin{equation}
 D(\alpha) = \frac{1}{b^2}(\pi \mu \gamma (b^2))^{( Q-2 \alpha )/b} 
  \frac{\gamma(2b \alpha - b^2)}
   {\gamma(2 - \frac{2 \alpha}{b} + \frac{1}{b^2})} ~.
\label{Da}
\end{equation}
The constraint does not determine the unique solution, however there is
a quite strong constraint that the quantities obtained should be related by
the duality $b \leftrightarrow 1/b$.
This duality should be understood by also replacing the cosmological
constant $\mu$ with $\tilde{\mu}$ satisfying
\begin{equation}
 \pi \tilde{\mu} \gamma\left(\frac{1}{b^2}\right) = 
  \left( \pi \mu \gamma (b^2)\right)^{1/b^2} ~.
\end{equation}
The general three point functions can be evaluated in the similar way
and their explicit forms are obtained in 
\cite{Dorn,zamozamo,Teschner1,Teschner2,Teschner3}.

%%%%%%%%%%%%%%%%%%%%%%%%%%%%%%%%%%%%%%%%%%%%%%%%%%%%%%%%%%%%%%%%%%%%

\section{One Point Function on a $\brp$}
\label{1pf}

For the oriented surface with boundary, there are several constraints of
the theory which are called as the sewing constraints \cite{sewingbs}, and 
for the unoriented surface, there are three types of additional
constraints \cite{sewingcs}.
In this section, we use the constraint related to the 
one point functions on the $\brp$
and in the next section we see
the other two types of constraints from M\"obius strip and Klein bottle
amplitudes.

The one point function on a $\brp$ can be calculated by using the mirror
technique. In the case of the one point function on a disk, 
we can use the upper half plane by conformal mapping from the disk.
Then, we can map from the upper half plane to the whole plane by using
the involution $I(z)=\bar{z}$. 
There is a fixed line $\im z = 0$, which corresponds to the boundary. 
In the case of the one point function on a $\brp$, 
we can also use the upper half plane, however we should use other involution 
$I(z) = -1/\bar{z}$ and there is no boundary. 
By using these mirror techniques, 
the one point function on a $\brp$ can be written as
\begin{equation}
 \langle V_{\alpha} (z,\bar{z}) \rangle{}_{\brp} = 
  \frac{U(\alpha)}{|1+z\bar{z}|^{2 \Delta_{\alpha}}} ~,
\end{equation} 
where the $z$ dependence is determined by the conformal symmetry.

In order to determine the coefficient $U(\alpha)$, we use the
two point function including the degenerate field $V_{-b/2}$
just like the bulk case as
\begin{equation}
 \langle V_{-\frac{b}{2}} (z,\bar{z}) V_{\alpha} (w,\bar{w}) 
  \rangle {}_{\brp} ~.
\end{equation}  
When two points $z$ and $w$ are close, it is natural to use the
OPE (\ref{OPE}) and we can write as
\begin{align}
 &\langle V_{-\frac{b}{2}} (z,\bar{z}) V_{\alpha} (w,\bar{w}) 
 \rangle {}_{\brp} =
 \frac{|1+w \bar{w}|^{2 \Delta_{\alpha} - 2 \Delta_{-b/2}}}
      {|1+z \bar{w}|^{4 \Delta_{\alpha}}} \times \nn & \qquad \qquad \times
  \left( C_+(\alpha) U \left(\alpha-\frac{b}{2} \right) {\cal F}_+ (\eta) +
   C_-(\alpha) U \left(\alpha+\frac{b}{2} \right) 
   {\cal F}_- (\eta)  \right) ~,
\label{2pf}  
\end{align}
where we define the cross ratio as
\begin{equation}
 \eta = \frac{|z-w|^2}{(1+z\bar{z})(1+w \bar{w})} ~.
\end{equation}
Since this correlation function includes the degenerate field 
$\Phi_{1,2}=V_{-b/2}$ which satisfies (\ref{diff}), 
the conformal blocks ${\cal F}_{\pm}$ also satisfy the differential
equation and they can be obtained by solving the differential equation. 
These solutions are expressed by the hypergeometric functions as
\begin{align}
 {\cal F}_+ (\eta) &= \eta^{b \alpha} (1 - \eta)^{b \alpha} 
  F(-1 + 2 b \alpha - 2 b^2, 2 b \alpha , 2 b \alpha - b^2 ; \eta) ~,\nn
 {\cal F}_- (\eta) &= \eta^{1 -b \alpha + b^2 } (1 - \eta)^{b \alpha}
  F(1+b^2, -b^2, 2 - 2 b \alpha +b^2 ; \eta) ~. 
\end{align}
Some properties of the hypergeometric functions are summarized in
appendix \ref{formulae}.

We should notice that the involution $I$ acts to the field as
\begin{equation}
 I : V_{\alpha} (z,\bar{z}) \to \epsilon_{\alpha} 
  V_{\alpha} \left( - \frac{1}{\bar{z}} , - \frac{1}{z} \right) ~,
\end{equation}
where the phase factor should be $\epsilon_{\alpha} = \pm 1$ since the
product of two involutions is the identity.
In the rational conformal field theory case, the label of fields takes
a discrete number, therefore we can choose an arbitrary sign for the
different fields as long as they are consistent with the OPE.
On the other hand, the label of fields in our case takes the continuous
value. Thus we can see that the consistency of OPE implies $\epsilon = +1$.
By using this fact, we find
\begin{equation}
 \langle V_{\alpha} (z,\bar{z}) \cdots \rangle =
 \left \langle V_{\alpha} \left( - \frac{1}{\bar{z}}, - \frac{1}{z}\right)
   \cdots \right\rangle ~,
\label{sign}
\end{equation}
and this equation gives constraint to the one point function
\cite{sewingcs}.

The following two point function can be calculated in the similar way as
\begin{align}
 &\left\langle V_{-\frac{b}{2}} (z,\bar{z}) 
 V_{\alpha} \left(- \frac{1}{\bar{w}},- \frac{1}{w}\right) 
  \right\rangle {}_{\brp} =
 \frac{|1+w \bar{w}|^{2 \Delta_{\alpha} - 2 \Delta_{-b/2}}}
      {|1+z \bar{w}|^{4 \Delta_{\alpha}}} \times \nn &  \qquad \qquad \times 
  \left( C_+(\alpha) U \left (\alpha-\frac{b}{2} \right) 
    {\cal F}_+ (1 - \eta) +
   C_-(\alpha) U \left(\alpha+\frac{b}{2} \right) 
    {\cal F}_- (1 - \eta)  \right) ~.
\label{2pfm}
\end{align}  
Using the properties of the hypergeometric functions in appendix
\ref{formulae}, we obtain
\begin{align}
 {\cal F}_+ (1-\eta) &= 
  \frac{\Gamma (2 b \alpha - b^2) \Gamma (1 - 2b\alpha + b^2)}
       {\Gamma (1+b^2) \Gamma (- b^2)} {\cal F}_+  (\eta)
  + \nn &  \qquad \qquad \qquad + 
  \frac{\Gamma (2 b \alpha - b^2) \Gamma (- 1 + 2b\alpha - b^2)}
       {\Gamma ( -1 + 2 b \alpha - 2 b^2 ) \Gamma ( 2 \alpha b)} 
   {\cal F}_- (\eta) ~, \nn
 {\cal F}_- (1-\eta) &= 
  \frac{\Gamma (2 - 2 b \alpha + b^2) \Gamma (1 - 2b\alpha + b^2)}
       {\Gamma (1 - 2 b \alpha) \Gamma (2 - 2 \alpha b + 2 b^2)} 
    {\cal F}_+ (\eta) + \nn & \qquad \qquad \qquad + 
  \frac{\Gamma (2 - 2 b \alpha + b^2) \Gamma (- 1 + 2b\alpha - b^2)}
       {\Gamma (1 + b^2) \Gamma ( -b^2)} {\cal F}_- (\eta) ~.
\end{align} 
Now we can compare two quantities (\ref{2pf}) and (\ref{2pfm}) by using
(\ref{sign}).
Then the following constraints are obtained as
\begin{align}
 U\left(\alpha - \frac{b}{2} \right)
 &= \frac{\Gamma (2 b \alpha - b^2) \Gamma (1 - 2 b \alpha + b^2)}
        {\Gamma (1+b^2) \Gamma (-b^2)}  
               U\left(\alpha - \frac{b}{2} \right) \nn
  & \qquad - \frac{\pi \mu} {\gamma (-b^2)} 
   \frac{\Gamma (1 - 2 b \alpha + b^2) \Gamma (-1 + 2 b \alpha - b^2)}
        {\Gamma (2 b \alpha) \Gamma (2 - 2 b \alpha + 2 b^2 )}  
               U\left(\alpha + \frac{b}{2} \right) ~,\nn
 U\left(\alpha + \frac{b}{2} \right)
 &= \frac{\Gamma (2 - 2 b \alpha + b^2) \Gamma (-1 +2 b \alpha - b^2)}
        {\Gamma (1+b^2) \Gamma (-b^2)}  
               U\left(\alpha + \frac{b}{2} \right) \nn
  & \qquad - \frac {\gamma (-b^2)} {\pi \mu}
   \frac{\Gamma (2- 2 b \alpha + b^2) \Gamma (2 b \alpha - b^2)}
        {\Gamma (1- 2 b \alpha) \Gamma (-1 + 2 b \alpha - 2 b^2 )}  
               U\left(\alpha - \frac{b}{2} \right) ~.
\label{constraint}
\end{align} 
These constraints can be solved and the solutions are of the forms as
\begin{align}
 U(\alpha) = \frac{2}{b} ( \pi \mu \gamma (b^2))^{(Q- 2 \alpha)/2b} 
  \Gamma (2 b \alpha - b^2) \Gamma \left( \frac{2 \alpha}{b} -
 \frac{1}{b^2} -1\right) f (\alpha) ~.
\end{align}
The function $f (\alpha)$ is given by the linear combination of the 
following two functions as
\begin{equation}
 \cos \left(  \left(b + \frac{1}{b} \right) \pi 
  \left( \alpha - \frac{Q}{2} \right)\right) ~,~~
 \cos \left(  \left(b - \frac{1}{b} \right) \pi 
  \left( \alpha - \frac{Q}{2} \right)\right) ~,
\label{1pfc}
\end{equation} 
where we have used the duality $b \leftrightarrow 1/b$ in order to restrict
the form of the solutions.
We should notice that these solutions satisfy the reflection relation
\cite{zamozamo}
\begin{equation}
 U(\alpha) = U(Q-\alpha)D(\alpha) ~,
\end{equation}
where $D(\alpha)$ is the coefficient of the two point function
(\ref{Da}).
Although we cannot determine the coefficients at this level,
they can be fixed by considering the M\"{o}bius strip amplitude as we
will see in the next section.

%%%%%%%%%%%%%%%%%%%%%%%%%%%%%%%%%%%%%%%%%%%%%%%%%%%%%%%%%%%%%%%%%%%%

\section{Crosscap State and Modular Bootstrap}
\label{crosscap}

It is convenient to introduce the boundary states and the crosscap
state for considering the Liouville field theory on the annulus,
M\"{o}bius strip and Klein bottle.
First, let us review the analysis of the boundary states \cite{FZZ,ZZ}.
The Virasoro character of the general non-degenerate representation 
$ \alpha = Q/2 + iP $ is given by
\begin{equation}
 \chi_{P} (\tau) = {\mbox{Tr}}_{{\cal H}_P} ( q^{ L_0 - \frac{c}{24}}) 
                 = \frac{q^{P^2}}{\eta (\tau)} ~,
\end{equation} 
where the eta function $\eta (\tau)$ is defined in appendix \ref{formulae}.
The modular transformation can be written as 
\begin{equation}
\chi_{P} \left(- \frac{1}{\tau} \right) = 
 \sqrt{2} \int d P' \chi_{P'} (\tau) e^{4 \pi i P P'} ~.
\end{equation}
For the degenerate state $\Phi_{m,n}$, the character is 
\begin{align}
 \chi_{m,n} (\tau) &= \frac{q^{- \frac{1}{4}\left(\frac{m}{b} + nb \right)^2 } 
               - q^{- \frac{1}{4}\left(\frac{m}{b} - nb \right)^2 }}
                   {\eta (\tau)} ~, 
\end{align}
which transforms under the modular transformation as
\begin{align}
 \chi_{m,n} \left( - \frac{1}{\tau} \right) 
 &= \sqrt{2} \int d P \chi_{P} (\tau) \left(
    \cosh \left( 2 \pi P \left(\frac{m}{b}  + n b \right)  \right)
 -    \cosh \left( 2 \pi P\left(\frac{m}{b}  - n b \right)  \right)
    \right)  \nn
 & = 2 \sqrt{2} \int d P \chi_{P} (\tau) 
  \sinh \left( \frac{2 \pi m P}{b} \right) \sinh (2 \pi n b P) ~.
\end{align}

The boundary states are described in terms of the Ishibashi states 
\cite{ishibashi} which satisfy
\begin{equation}
 {}_I \bra{P} q^{\frac{1}{2} \left( L_0 + \bar{L_0} - \frac{c}{12} \right)}
 \ket{P'}{}_I  = \delta_{P,P'}\chi_{P} (\tau) ~. 
\end{equation}
The general boundary states can be written by the linear combination of
the Ishibashi states.
The coefficients correspond to the one point functions since
they can be calculated by the overlaps between the boundary
states and closed string states.

The one point function on a pseudosphere was obtained in \cite{ZZ} and
the corresponding boundary states are labeled by $(m,n)$ as
\begin{equation}
{}_C \bra{m,n} = \int d P \Psi_{m,n} (P) {}_I\bra{P} ~.  
\label{mn}
\end{equation}
The boundary state $\ket{1,1}_C$ can be interpreted as a basic state and
the wave function $\Psi_{1,1} (P)$ is 
\begin{equation}
 \Psi_{1,1} (P) = \frac{2^{3/4} 2 \pi i P}
    { \Gamma (1 - 2 i b P) \Gamma \left(1 - \frac{2 i P}{b}\right) }
      (\pi \mu \gamma (b^2))^{ - i P /b} ~.
\end{equation}
The other wave functions $\Psi_{m,n}$ are expressed in this basis as
\begin{equation}
 \Psi_{m,n} (P) = \Psi_{1,1} (P) 
 \frac{ \sinh \left( \frac{2 \pi m P}{b}\right) \sinh (2 \pi n b P)}
      { \sinh (\frac{2 \pi P}{b}) \sinh (2 \pi b P)} ~.
\end{equation} 

There are the other kind of boundary states which correspond to the one
point functions on the disk \cite{FZZ}.
The wave functions can be labeled by a continuous number $s$ and they
are given by
\begin{equation}
 \Psi_s (P) = \frac{2^{-1/4} \Gamma (1 + 2 i b P) 
   \Gamma \left( 1 + \frac{ 2 i P}{b} \right) \cos (2 \pi s P)}
    { - 2 i \pi P} ( \pi \mu \gamma (b^2))^{- i P /b} ~.
\label{s}
\end{equation}
The normalization of wave functions is determined by using the basic
boundary states.

Next, we construct the crosscap state. 
For the M\"{o}bius strip amplitudes, it is convenient to introduce the
following characters \cite{sagnotti1} as
\begin{equation}
 \hat{\chi}_{\alpha} (q) = e^{- \pi i (\Delta_{\alpha} - \frac{c}{24})} 
  \chi_{\alpha} ( - \sqrt{q}) ~.
\label{hatchi}
\end{equation}
The modular transformation of the M\"{o}bius strip can be performed by
so called $P$ matrix ($P=\sqrt{T} S T^2 S \sqrt{T}$). This matrix
transforms $\tau \to -1/(4 \tau)$ and for the character of the
non-degenerate representation it can be given by
\begin{equation}
 \frac{e^{2 \pi i \left(- \frac{1}{4 \tau} \right) P^2}}
      {\eta\left(- \frac{1}{4 \tau}\right)} = 
 \int d P'  e^{2 \pi i P P'} \frac{e^{2 \pi i \tau {P '}^2}}{\eta (\tau)} ~,
\label{PforC}
\end{equation}
and for the character of the degenerate representation it can be written
as\footnote{In the previous version, there was a sign mistake in the second
term of the first equation.  
I am grateful to S.~Hirano and Y.~Nakayama for pointing
out this error.}
\begin{align}
 & \frac{e^{-2 \pi i \left(- \frac{1}{4 \tau}\right) 
  \frac{1}{4}\left( \frac{m}{b} + n b\right)^2 } - (-1)^{mn}
 e^{-2 \pi i \left(- \frac{1}{4 \tau}\right) 
  \frac{1}{4}\left( \frac{m}{b} - n b\right)^2 } }
           { \eta \left(-\frac{1}{4 \tau}\right)} = \nn 
 & \qquad = \int d P \frac{e^{2 \pi i \tau {P }^2}}{\eta (\tau)} 
  \left( \cosh \left( \pi P \left( \frac{m}{b} + nb\right)\right)
   - (-1)^{mn} 
  \cosh\left( \pi P \left( \frac{m}{b} - nb\right)\right)\right) ~.
% & \qquad \qquad  =  2 \int d P \frac{e^{2 \pi i \tau {P }^2}}{\eta (\tau)} 
%         \sinh \left( \frac{\pi m P}{b}\right) \sinh ( \pi n b P) ~.
\label{PforD}
\end{align}

In the case of crosscap state, the Ishibashi states are defined by
\begin{align}
 & {}_I \bra{C,P} 
  q^{\frac{1}{2}\left( L_0 + \bar{L}_0 - \frac{c}{12} \right)}
       \ket{C,P'}_I = \delta_{P,P'}\chi_P (\tau) ~, \nn
 & {}_I \bra{B,P} 
  q^{\frac{1}{2}\left( L_0 + \bar{L}_0 - \frac{c}{12} \right)}
       \ket{C,P'}_I = \delta_{P,P'}\hat{\chi}_P (\tau) ~.
\end{align}
In this basis, the crosscap state is represented as
\begin{equation}
 {}_C \bra{C} = \int d P \Psi_C (P) {}_I \bra{C,P} ~,
\label{C}
\end{equation}
where $\Psi_C (P)$ is the wave function corresponding to the crosscap state.

In order to determine the wave function $\Psi_C (P)$, we use the character
of  the identity representation $(m,n)=(1,1)$.
The modular transformation is given in (\ref{PforD}) and it can be 
interpreted as
\begin{equation}
 \hat{\chi}_{1,1} \left(- \frac{1}{\tau}\right) 
  = \int d P \hat{\chi}_{P} (\tau) \Psi_{1,1} (P) \Psi_C  (-P)  ~.
\end{equation}
This equation determines the wave function including the
normalization factor as
\begin{align}
\Psi_C (P) &= \frac{ 2^{-3/4} \Gamma (1 + 2 i b P) 
 \Gamma (1 + \frac{2 i P}{b})}{- 2 i \pi P} ( \pi \mu \gamma
 (b^2))^{-iP/b}  \times \nn 
 & \qquad
  \times \left( \cosh \left(  \pi P \left( b + \frac{1}{b}\right) \right)
  +  \cosh \left(  \pi P \left( b - \frac{1}{b}\right) \right)
  \right) ~.
\label{cs}
\end{align}
This also determines the precise form of the one point function on the
$\brp$ (\ref{1pfc}).

Because we obtain the precise form of the crosscap state, we can
calculate the other partition functions straightforwardly.
The overlaps between the boundary states $\ket{m,n}_C$
(\ref{mn}) and the crosscap state $\ket{C}_C$ are given by
\begin{align}
 Z_{m,n} (\tau) & = \int d P \hat{\chi}_{P} (\tau)\Psi_{m,n} (P) \Psi_C (-P)  
   \nn & = 2 \int d P \hat{\chi}_{P} (\tau)
  \frac{ \sinh \left( \frac{2 \pi m P}{b} \right) \sinh (2 \pi n b P)
    \cosh \left( \frac{ \pi P}{b} \right) \cosh (\pi  b P)
    }{\sinh \left( \frac{2 \pi P}{b} \right) \sinh (2 \pi b P) } ~.
\end{align}
By using the formula
\begin{equation}
 \frac{\sinh (2 \pi n b P) \cosh (\pi b P)}{\sinh (2 \pi b P)} = 
  \sum_{l=0,1,\cdots}^{n-1} \cosh( \pi b P (2l+1)) ~,
\end{equation}
we find
\begin{equation}
  Z_{m,n} (\tau) = \sum_{k=0,1,\cdots}^{m-1} \sum_{l=0,1,\cdots}^{n-1}
    \hat{\chi}_{2k+1,2l+1} \left( -\frac{1}{\tau}\right)~.
\end{equation}
We should note that the coefficient of the character of the identity
representation is less than one, which means that the crosscap state we
have constructed is the irreducible one.

The other type of the M\"{o}bius strip amplitudes correspond to the overlaps
between the boundary states parametrized by $s$ and the crosscap state
as
\begin{align}
 Z_s (\tau) &= \int d P \hat{\chi}_P (\tau) \Psi_s (P) \Psi_C (-P) \nn 
     & = \int d P d P' \hat{\chi}_{P'} \left( - \frac{1}{\tau}\right) 
       e^{ 2 \pi i P P'} \Psi_s (P) \Psi_C (-P) \nn
     & = \int d P'  \hat{\chi}_{P'} \left( - \frac{1}{\tau}\right)\rho (P')~, 
\end{align}
where $\rho (P') $ is the density of states.
The last one is the Klein bottle amplitude, which is given by
\begin{align}
 Z (\tau) &= \int d P \chi_P (\tau) \Psi_C (P) \Psi_C (-P) \nn 
     & = \int d P d P' \chi_{P'} \left( - \frac{1}{\tau}\right) 
        e^{ 4 \pi i P P'}\Psi_C (P) \Psi_C (-P) \nn
     & = \int d P'\chi_{P'} \left( - \frac{1}{\tau}\right) \rho (P') ~.
\end{align}
In the case of the boundary states, the density of states can be
calculated by the other method and we can compare them.
It is interesting to compare these densities of states with the ones
obtained by other methods if we could 
also in the case of crosscap state.

%%%%%%%%%%%%%%%%%%%%%%%%%%%%%%%%%%%%%%%%%%%%%%%%%%%%%%%%%%%%%%%%%%%%%

\section{Conclusion}
\label{conclusion}

Liouville field theory on an unoriented surface is investigated.
The basic information is given by the one point function on a 
$\brp$.
Since it is difficult to calculate in general, we use the trick which
was developed for the bulk three point function \cite{Teschner1} and for
the one point function on a disk \cite{FZZ} and on a pseudosphere
\cite{ZZ}.
The degenerate states satisfy some differential equations, and hence the
two point functions including these states are calculable.
By assuming the crossing symmetry, we obtain the constraint for the
general one point function (\ref{constraint}).
Although there are plenty of solutions of the constraint, 
we can choose one of them (\ref{cs})
by making use of the modular bootstrap.

Since Liouville field theory is a typical example of the non-rational
conformal field theory, the application to the other backgrounds, 
e.g., $AdS_3$ spaces \cite{AdSoplane}, 
might be done by using the methods we have used.
Apart from the solvable property, Liouville field theory is interesting
because it can be embedded into the full superstring theory.
For that purpose, we should extend our analysis to the supersymmetric
case like \cite{FH,ARS}.
If we can apply to the consistent superstring theory, the
orientifolds in a non-trivial background can be constructed and we
may see interesting phenomena in the system with the non-trivial
orientifolds.

%%%%%%%%%%%%%%%%%%%%%%%%%%%%%%%%%%%%%%%%%%%%%%%%%%%%%%%%%%%%%%%%%%%%%

\subsection*{Acknowledgement}

We would like to thank K. Hosomichi for useful discussions.

%%%%%%%%%%%%%%%%%%%%%%%%%%%%%%%%%%%%%%%%%%%%%%%%%%%%%%%%%%%%%%%%%%%%%%

\appendix

\section{Several Useful Formulae}
\label{formulae}

The hypergeometric functions have the following properties under the
reparametrizations 
\begin{align}
 F(a,b,c;\eta)&= (1-\eta)^{c-a-b} F (c-a,c-b,c;\eta) ~,\\
 F(a,b,c;1-\eta)&= \frac{\Gamma (c) \Gamma (c-a-b)}{\Gamma (c-a) \Gamma (c-b)}
  F (a,b,1-c+a+b;\eta) \nn
   &+ \eta^{c-a-b}   \frac{\Gamma (c) \Gamma (a+b-c)}{\Gamma (a) \Gamma (b)}
        F (c-a, c-b, 1+c-a-b;\eta) ~.
\end{align}

We often use the following formulae for Gamma function as
\begin{align}
 &\Gamma (1+z) = z \Gamma (z) ~,\\
 &\Gamma (1-z) \Gamma (z) = \frac{\pi}{\sin(\pi z)}  ~,\\
 &\Gamma (1 + ix) \Gamma (1 - ix)  = \frac{\pi x}{\sinh (\pi x)}~,
\end{align}
where $z$ is an arbitrary complex number and $x$ is a real number.

The Dedekind eta function is defined by
\begin{equation}
 \eta (\tau) = q^{\frac{1}{24}} \prod_{n=1}^{\infty} (1 - q^n) ~,
\end{equation}
where $q=\exp (2 \pi i \tau)$ and its modular transformation is given by
\begin{equation}
 \eta (\tau + 1) =
  e^{\pi i / 12} \eta (\tau) ~,~~
 \eta \left(- \frac{1}{\tau}\right) =
  \sqrt{-i \tau} \eta (\tau) ~.
\label{etam}
\end{equation} 

%%%%%%%%%%%%%%%%%%%%%%%%%%%%%%%%%%%%%%%%%%%%%%%%%%%%%%%%%%%%%%%%%%%%%%%%%

\end{document}